\documentstyle[12pt]{article}
 \begin{document}

 \begin{center}
 {\large{\bf Non-abelian self-duality from self-interaction}}\\
 \vskip 1cm
 {\bf A. Khoudeir}\\
 {\ Instituto de F\'{\i}sica, Universidad Nacional Aut\'onoma de M\'exico\\
 Apdo. Postal 20-364, 01000 M\'exico D. F. M\'exico}\\

 and

 {\it Centro de Astrof\'{\i}sica Te\'orica, Departamento de F\'{\i}sica,
 Facultad de Ciencias, Universidad de los Andes,\\
 M\'erida, 5101,Venezuela.}\\

 \end{center}

 \begin{abstract}

 The non-abelian self-dual action in three dimensions is derived using 
 the self-interaction mechanism.

 \end{abstract}

 Self-duality in three dimensions was proposed initially 
 by Townsend et. al. \cite{tpvn}
 as an alternative to the topologically massive theory\cite{djt}. 
 In principle, they seem different descriptions of a locally 
 massive spin 1 physical excitation: the self-dual theory is 
 described by a non-gauge invariant first order action while the 
 topologically massive action is written down in a gauge invariant 
 second order formulation. Both actions have an abelian 
 Chern-Simons term ($\epsilon^{mnp}A_{m}\partial_{n}A_{p}$). 
 Despite these differences, Deser and Jackiw stablished 
 that both theories are locally equivalent through the existence of a 
 master action, even in the presence of external sources\cite{dj}. 
 Moreover, both theories 
 are dual equivalent\cite{jste} and the self-dual theory can be seen as a 
 gauged fixed version of the topologically massive theory\cite{rjap}. 
 The self-dual theory for gravity and for higher spin in three dimensions 
 was achieved in \cite{ak1} and \cite{ak2}, respectively.
 If glogal properties are considered, the equivalence is modified, 
 for instance, the partition functions of the self dual and topologically 
 massive theories are not the same but they are related in the following 
 way: $Z_{SD} = Z_{CS}Z_{TM}$\cite{pial} (where $Z_{CS}$ is the partition 
 function of the abelian Chern-Simons action).

 The non-abelian generalization of the topologically massive theory was 
 given in \cite{djt} while the non-abelian self-dual theory was formulated 
 independently by McKeon \cite{mck} and Arias, et. al.\cite{pr2}, which has a 
 structure of a Freedman-Townsend action\cite{ft}. 

 In this letter, starting from an appropiate master action, we will derive the 
 non-abelian self-dual action using the self-interaction mechanism\cite{des}.

 We will start by considering the following master action\cite{ak}
 \begin{equation}
 I = \int d^3 x [-\mu\epsilon^{mnp}A_{m}\partial_n a_p  - 
 \frac{1}{2}\mu^{2}a_m a^m 
 - \mu\epsilon^{mnp}A_m \partial_n v_p + \frac{1}{2}\mu\epsilon^{mnp}v_{m}
 \partial_{n}v_{p} ]
 \end{equation}
 This action can be seen as the coupling between a Maxwell field ($A_m$) 
 and a vector field ($v_{m}$) described by an abelian Chern-Simons action 
 through a three dimensional BF topological term.
 Independent variations in the $a_m$, $v_m$ and $A_m$ fields, 
 yield the following equations of motion
 \begin{equation}
 a^m = -\frac{1}{2}\mu \epsilon^{mnp}f_{np(A)}, 
 \end{equation}
 \begin{equation}
 \epsilon^{mnp}\partial_{n}[A_p - v_p ] = 0
 \end{equation}
 and
 \begin{equation}
 \epsilon^{mnp}\partial_{n}[a_p + v_p ] = 0,
 \end{equation}
 where $f_{mn(A)} = \partial_{m}A_{n} - \partial_{n}A_{m}$. 
 The last two equations can be solved locally. We have
 \begin{equation}
 v_m = A_m + \partial_m \phi
 \end{equation}
 and 
 \begin{equation}
 a_m = -v_m + \partial_m\sigma . 
 \end{equation}
 The master action has abelian gauge invariance
 \begin{equation}
 \delta A_{m} = \partial_{m}\lambda_1 \quad \delta v_{m} = \partial_{m}\lambda_2
 \end{equation}
 Substituting the equations (2) and (5), into the master 
 action lead to the action for the abelian topologically massive theory 
 \begin{equation}
 I = \int d^3 x [ -\frac{1}{4}f^{mn}_{(A)}f_{mn(A)} - 
 \frac{1}{4}\mu\epsilon^{mnp}A_m f_{np(A)} ].
 \end{equation}
 On the other hand, we can eliminate the $a_{m}$ and $A_{m}$ fields, through 
 the use of equations (5) and (6) in order to obtain
 \begin{equation}
 I = \int d^3 x [-\frac{1}{2}\mu^2 (v_m -\partial_m\phi )
 (v^m  -\partial^m\phi) + 
 \frac{1}{2}\mu\epsilon^{mnp}v_{m}\partial_{n}v_{p} ],
 \end{equation}
 which is invariant under the following abelian gauge transformations
 \begin{equation}
 \delta v_{m} = \partial_{m}\lambda_1 , \quad \delta\phi = \lambda_1 .
 \end{equation}
 Fixing the gauge $\phi = 0$, we obtain the non-gauge invariant 
 self-dual action. Then, the proposed 
 master action show the equivalence (at classical level) between the 
topologically and
 self-dual theories. 
 The master action that we are considering is locally equivalent to the master 
action
 of Deser and Jackiw, as can be seen after eliminating only the $v_m$ 
 field and is written down as
 \begin{equation}
 I = \int d^3 x [-\mu\epsilon^{mnp}A_{m}\partial_n a_p  - 
 \frac{1}{2}\mu^{2}a_m a^m 
 - \frac{1}{2}\mu\epsilon^{mnp}A_{m}\partial_{n}A_{p} ]
 \end{equation} 
 Introducing the Lie-algebra valued vectors $A_m = A_{m}^{i}T^{i}$, 
 $a_m = a_{m}^{i}T^{i}$ and the Lie-algebra valued field strength 
 $F_{mn} = F_{mn}^{i}T^{i}$, where the generators $T^{i}$ of 
 the gauge group are normalized by $T^{i}T^{j} = \delta^{ij}$,
 the non-abelian generalization of the master action of 
 Deser and Jackiw obtained by replacing 
 ordinary derivative by covariant derivative, $f_{mn} = \partial_{m}A_{n} - 
 \partial_{n}A_{m}$ $\rightarrow$ $F_{mn} = \partial_{m}A_{n} - 
 \partial_{n}A_{m} + [A_{m},A_{n}]$ and considering the non-abelian 
 Chern-Simons term is 
 \begin{equation}
 I = \mu tr\int d^3 x [\epsilon^{mnp}a_{m}F_{np}  - 
 \frac{1}{2}\mu a_m a^m 
 - \frac{1}{2}\epsilon^{mnp}A_{m}(\partial_{n}A_{p} +  \frac{2}{3}A_{n}A_{p})]
 \end{equation} 
 and only can reproduce the non-abelian version of the topologically massive 
 theory after eliminating the $a_{m}$ field by using its equation of motion
 ($a^{m} = \epsilon^{mnp}F_{np}$). 
 On the other hand, the equation of motion obtained by independent variations in
 $A_{m}$ has no known solutions and in consecuence the non-abelian master 
 action of Deser and Jackiw can not reproduce the non-abelian self-dual 
 action. The non-abelian topologically massive 
 theory can be deduced from the self-interaction mechanism\cite{aa}.

 Now, we will consider for simplicity a triplet of $SU(2)$  free vector 
 fields $A_{m}^{i}$ coupled with a triplet of $SU(2)$ free vector fields 
 $v_{m}^{i}$ ($i = 1,2,3$). The action is 
 \begin{equation}
 I_o = \int d^3 x [-\mu\epsilon^{mnp}A_{m}^{i}\partial_n a_p^{i}  - 
 \frac{1}{2}\mu^{2}a_m^{i} a^{mi} 
 - \mu\epsilon^{mnp}A_m^{i} \partial_n v_p^{i} + 
\frac{1}{2}\mu\epsilon^{mnp}v_{m}^{i}
 \partial_{n}v_{p}^{i} ] .
 \end{equation}
 This action has two global simmetries. One is the global $SU(2)$ symmetry
 \begin{equation}
 \delta_{\omega}X = g\epsilon^{ijk}X^{j}\omega^{k}
 \end{equation}
 where X = (A, a, v) and the other global symmetry is given by
 \begin{equation}
 \delta_{\rho}A_{m}^{i} = g\epsilon^{ijk}[a_{m}^{j} + v_{m}^{j}]\rho^{k}; \quad 
 \delta_{\rho}a_{m}^{i} = 0 = \delta_{\rho}v_{m}^{i}. 
 \end{equation} 
 Under these transformations, the action changes by a total derivative. 
   
 The Noether currents associated with the global symmetries are 
 \begin{equation}
 j^{mi} = -\mu g\epsilon^{mnp}\epsilon^{ijk}A_{n}^{j}[a_{p}^{k} + v_{p}^{k}] 
 + \frac{1}{2}\mu g\epsilon^{mnp}\epsilon^{ijk}v_{n}^{j}v_{p}^{k} 
 \end{equation}
 and
 \begin{equation}
 K^{mi} = -\frac{1}{2}\mu g\epsilon^{mnp}\epsilon^{ijk}[a_{n}^{j} + v_{n}^{j}]
 [a_{p}^{k} + v_{p}^{k}] . 
 \end{equation}
 These currents are conserved on-shell. Now, we will couple these 
 Noether currents to the action $I_0$ through the corresponding self-interaction 
term
 defined by 
 \begin{equation}
 j^{mi} \equiv \frac{\delta I_{SI}}{\delta v_{m}^{i}}, \quad 
 K^{mi} \equiv \frac{\delta I_{SI}}{\delta A_{m}^{i}} . 
 \end{equation}
 We find
 \begin{eqnarray}
 I_{SI} &=& g \mu \int d^{3}x [-\epsilon^{mnp}\epsilon^{ijk}v_{m}^{i}A_{n}^{j}
 a_{p}^{k} -\frac{1}{2}\epsilon^{mnp}\epsilon^{ijk}v_{m}^{i}v_{n}^{j}
 A_{p}^{k} \\ \nonumber
 &-& \frac{1}{2}\epsilon^{mnp}\epsilon^{ijk}A_{m}^{i}a_{n}^{j}a_{p}^{k}
 + \frac{1}{6}\epsilon^{mnp}\epsilon^{ijk}v_{m}^{i}v_{n}^{j}v_{p}^{k}] .
 \end{eqnarray}
 The self-interaction mechanism stops here since no other derivative terms 
 appear in $I_{SI}$. Now, we add $I_{SI}$ to $I_{o}$. The last term in eq. (13) 
 combines with the last term in eq. (19) to give a Chern-Simons term for 
 the $v_m$ field. The non-abelian action is
 \begin{eqnarray}
 I &=& \frac{1}{2}\mu\int d^{3}x[-\epsilon^{mnp}A_{m}^{i}(F_{np(a)}^{i} + 
 F_{np(v)}^{i} + 2g\epsilon^{ijk}a_{n}v_{p}^{k}) - \mu a_{m}^{i}
 a^{mi}\\ \nonumber 
 &+& \epsilon^{mnp}v_{m}^{i}(\partial_{n}v_{p}^{i} + \frac{1}{3}
 \epsilon^{ijk}v_{n}^{j}v_{p}^{k})] ,
 \end{eqnarray}
 or
 \begin{equation}
 I = \frac{1}{2}\mu\int d^{3}x[-\epsilon^{mnp}A_{m}^{i}F_{np(a + v)}^{i} 
 - \mu a_{m}^{i}a^{mi} 
 + \epsilon^{mnp}v_{m}^{i}(\partial_{n}v_{p}^{i} + \frac{1}{3}
 \epsilon^{ijk}v_{n}^{j}v_{p}^{k}) ] ,
 \end{equation}
 where
 \begin{equation}
 F_{mn(a)}^{i} = \partial_{m}a_{n}^{i} - \partial_{n}a_{m}^{i} + 
 g\epsilon^{ijk}a_{m}^{j}a_{n}^{k}
 \end{equation}
 and 
 \begin{equation}
 F_{mn(v)}^{i} = \partial_{m}v_{n}^{i} - \partial_{n}v_{m}^{i} + 
 g\epsilon^{ijk}v_{m}^{j}v_{n}^{k}
 \end{equation}
 are the field strengths for the $a_{m}^{i}$ and $v_{m}^{i}$ fields. 
 The self-interaction process combines the abelian gauge transformations 
 with the global ones giving rise to the following non-abelian local gauge 
 transformations
 \begin{eqnarray}
 \delta A_{m}^{i} &=& g\epsilon^{ijk}A_{m}^{j}\alpha^{k} ; \quad 
 \delta a_{m}^{i} = g\epsilon^{ijk}a_{m}^{j}\alpha^{k}\\ \nonumber
 \delta v_{m}^{i} &=& \partial_m\alpha^{i} + g\epsilon^{ijk}v_{m}^{j}\alpha^{k} 
 \end{eqnarray}
 and
 \begin{eqnarray}
 \delta A_{m}^{i} &=& \partial_{m}\kappa^{i} + 
 g\epsilon^{ijk}[a_{m}^{j} + v_{m}^{j}]\kappa^{k}\\ \nonumber
 \delta a_{m}^{i} &=& 0 = \delta v_{m}^{i}
 \end{eqnarray}
 Defining $\omega_{m} \equiv a_{m} + v_{m}$, the action is 
 rewritten down as
 \begin{eqnarray}
 I &=& \frac{1}{2}\frac{\mu}{g^{2}} tr\int 
d^{3}x[-\epsilon^{mnp}A_{m}F_{np(\omega)} 
 - \mu  (v_{m} - \omega_{m})(v^{m} - \omega^{m}) \\ \nonumber 
 &+& \epsilon^{mnp}v_{m}[\partial_{n}v_{p} + \frac{2}{3}v_{n}v_{p}] .
 \end{eqnarray} 
 This action was interpreted as the interaction between a Chern-Simons 
 and a BF($\epsilon AF$) topological terms propagating a massive spin 1 physical
 mode\cite{pr2}. Like as in the non-abelian topologically massive theory, 
invariance
 in the functional integral implies the quantization condition: 
$4\pi\frac{\mu}{g^2} =
 integer$. 

 We observe that $A_{m}$ play the role of a Lagrange multiplier. 
 Its equation of motion is 
 \begin{equation}
 F_{mn(\omega)} = 0
 \end{equation}
 which tell us that $\omega$ is a pure gauge.
 \begin{equation}
 \omega_{m} = U^{-1}\partial_{m} U .
 \end{equation}
 Then, the action becomes
 \begin{equation}
 I = \frac{1}{2} \frac{\mu}{g^{2}} tr\int d^{3}x [-\mu(v_{m} - 
U^{-1}\partial_{m} U)
 (v^{m} - U^{-1}\partial^{m} U) + \epsilon^{mnp}v_{m}(\partial_{n}v_{p} + 
 \frac{2}{3}v_{n}v_{p})] ,
 \end{equation}
 where the $v_{m}$ field appear coupled with a Stuckelberg field.
 Now, we have invariance under the following (finite) gauge transformations 
 \begin{equation}
 v_{m} \rightarrow g^{-1}\partial_{m}\partial_{m}g + g^{-1}v_{m}g , \quad 
 U \rightarrow Ug .
 \end{equation}
 This gauge invariance allow us to fix the gauge $U =1$, in order to obtain the 
 standard action for the non-abelian self-dual field $v_{m}$
 \begin{equation}
 I = \frac{1}{2}\frac{\mu}{g^{2}}  tr\int d^{3} [-\mu v_{m}v^{m} + 
\epsilon^{mnp}v_{m}
 (\partial_{n}v_{p} + \frac{2}{3}v_{n}v_{p})] .
 \end{equation}
 To conclude, we have derived the non-abelian self-dual action in 
 three dimensions using the self-interaction mechanism. Recently, a 
 dual version of a pure non-abelian Chern-Simons action was formulated
 \cite{comp}. It would be interesting to analyse the duality properties of 
 the self-dual and topologically masive theories at non-abelian level.

 \begin{center}
 {ACKNOWLEDGEMENTS}
 \end{center}

 The author would like to thank to Marti Ruiz Altaba for his hospitality 
 at Instituto de F\'{\i}sica de la Universidad Nacional Aut\'onoma de 
 M\'exico. Also, the author thanks Conicit-Venezuela for financial support.

 \end{document}